\documentclass[twocolumn,astrosymb]{aastex7}
\usepackage{amsmath}

\begin{document}

\title{Testing the Origin of Hot Jupiters with Atmospheric Surveys}

\author[orcid=0009-0005-9175-4768,gname=Lina,sname=D'Aoust]{Lina D'Aoust}
\affiliation{Department of Physics \& Astronomy, University of Waterloo, Waterloo, Ontario, N2L 3G1, Canada}
\affiliation{Department of Physics \& Astronomy, University of Victoria, Victoria, British Columbia, V8P 5C2, Canada}
\affiliation{Trottier Institute for Research on Exoplanets (IREx), Universit\'e de Montr\'eal, Canada}
\email[show]{linatheresed@gmail.com}  

\author[orcid=0009-0000-2192-6003,gname=Ben, sname=Coull-Neveu]{Ben Coull-Neveu} 
\affiliation{Department of Physics and Trottier Space Institute, McGill University, 3600 rue University, H3A 2T8 Montreal QC, Canada}
\email[show]{benjamin.coull-neveu@mail.mcgill.ca\\}

\author[orcid=0000-0002-1228-9820,gname=Eve,sname=Lee]{Eve J.~Lee}
\affiliation{Department of Astronomy \& Astrophysics, University of California, San Diego, La Jolla, CA 92093-0424, USA}
\affiliation{Department of Physics and Trottier Space Institute, McGill University, 3600 rue University, H3A 2T8 Montreal QC, Canada}
\affiliation{Trottier Institute for Research on Exoplanets (IREx), Universit\'e de Montr\'eal, Canada}
\email[show]{evelee@ucsd.edu}

\author[orcid=0000-0001-6129-5699,gname=Nicolas,sname=Cowan]{Nicolas B.~Cowan}
\affiliation{Department of Physics and Trottier Space Institute, McGill University, 3600 rue University, H3A 2T8 Montreal QC, Canada}
\affiliation{Department of Earth \& Planetary Sciences, 3450 University Street
Montreal, QC H3A 0E8, Canada}
\affiliation{Trottier Institute for Research on Exoplanets (IREx), Universit\'e de Montr\'eal, Canada}
\email[show]{nicolas.cowan@mcgill.ca}

\begin{abstract}

In spite of their long detection history, the origin of hot Jupiters remains to be resolved. While multiple dynamical evidence suggests high-eccentricity migration is most likely, conflicts remain when considering hot Jupiters as a population in the context of warm and cold Jupiters. Here, we turn to atmospheric signatures as an alternative mean to test the origin theory of hot Jupiters, focusing on population level trends that arise from post-formation pollution, motivated by the upcoming Ariel space mission whose goal is to deliver a uniform sample of exoplanet atmospheric constraints. We experiment with post-formation pollution by planetesimal accretion, pebble accretion, and disk-induced migration and find that an observable signature of post-formation pollution is only possible under pebble accretion in metal-heavy disks. If most hot Jupiters arrive at their present orbit by high-eccentricity migration while warm Jupiters emerge largely in situ, we expect the atmospheric water abundance of hot Jupiters to be significantly elevated compared to warm Jupiters. We report on the detectability of such signatures and further provide suggestions for future comparative atmospheric characterization between hot Jupiters and wide-orbit directly imaged planets to elucidate the properties of the dust substructures in protoplanetary disks. 

\end{abstract}

\section{Introduction} 

Hot Jupiters are one of the first kind of exoplanet ever detected. Yet, their origin remains debated. Broadly, three origin hypotheses have been proposed in the literature, including in situ formation, disk-induced migration, and high-eccentricity migration, all of which have their successes and shortcomings \citep[see][for a review]{Dawson2018}.
Among these three, in situ formation \citep[e.g.,][]{Batygin16} is the least likely given the need to coagulate cores of mass $\gtrsim 10M_\oplus$ \citep[e.g.,][]{Rafikov2006AtmospheresInstability,Lee2014,Piso2015MinimumOpacities,Savignac24} which is prohibitively large at such short orbital distances by e.g., pebble accretion \citep[e.g.,][]{Bitsch2018,Fung2018}. While continued planetesimal accretion with enough solids in the background disk could generate $\sim$10$M_\oplus$ cores on timescales $\lesssim$10$^4$ yr at orbital periods of a few days \citep[see e.g.,][their Section 1.1]{Lee2014}, observations find that the vast majority of $\lesssim$10$M_\oplus$ planets at these short orbital periods stay as sub-Neptunes rather than blowing up into hot Jupiters in situ \citep[e.g.,][]{Fressin13,Dong13,Petigura18}.

Between disk-induced and high-eccentricity migration, there is growing evidence that the latter is more likely, including the distribution of obliquities and eccentricities \citep{Petrovich2016WarmInteractions,Rice22}, the shape of the upper edge of the sub-Jovian desert \citep{Matsakos2016OnPlane,Owen2018PhotoevaporationDesert}, and the systematically older ages of true hot Jupiters compared to puffy sub-Neptunes \citep{Karalis25}. 
Constructing a single theory that explains hot Jupiters in the context of warm and cold Jupiters has been challenging, however. 
For example, approximately half of warm Jupiters are accompanied by sub-Neptune companions \citep{Huang2016WarmNeighbours}, a configuration that is hard to reconcile with high-eccentricity migration scenarios, which often lead to orbital instability \citep{Antonini2016DynamicalFriends}. Smaller obliquities of tidally detached warm Jupiter systems further support their more dynamically quiescent origin compared to hot Jupiters \citep[e.g.,][]{Rice22,Wang24}.
In addition, disk-induced migration can explain well the overall shape of the orbital period distribution of hot, warm and cold Jupiters but underestimates the typical mass of hot Jupiters \citep{Hallatt2020}. 

An alternative path to resolving the origin of hot Jupiters is to use chemical abundances. Early and recent studies have focused on the expected atmospheric composition of hot Jupiters depending on their initial formation locations \citep[e.g.,][]{Madhu14,Schneider21,Penzlin24}, based on the radial variation in chemical content in protoplanetary disks \citep[e.g.,][]{Oberg11,Molliere22}. These studies generally expect high C/O ratio as a signature of formation beyond the iceline as the water molecules are expected to be in solid form and gas giants, by definition, are mostly gas (i.e., gas-phase C/O is expected to be high so it follows that gas-dominated accretion leads to high C/O). Pollution of the envelope by continued solid accretion after (or during) formation can complicate this picture, however, as the infalling solids are expected to evaporate within the envelope \citep[e.g.,][]{Mousis09,Fortney13,Mordasini16,Shibata20,Vlahos24}. Support for post-formation pollution of gas giants may be found in the high atmospheric metallicities of our own Jupiter \citep[e.g.,][]{Oberg19} and HR 8799 planets (\citealt{Nasedkin24}; J-B.~Ruffio et al.~submitted), as well as low C/O ratio of $\beta$ Pic b \citep{GRAVITY-2020}. At short orbital periods, supersolar metallicity is reported for WASP-15b which is a spin-orbit misalignd hot Jupiter \citep{Kirk25}, and interior structure models infer high (superstellar) bulk metallicity of warm Jupiters \citep{Thorngren16}.

While the collection of dynamical evidence suggests distinct origins of hot vs.~warm Jupiters, a clear alternative test from atmospheric studies remains to be proposed. In this paper, we aim to develop population-level atmospheric signatures (i.e., trends) as a probe of disk-induced vs.~high-eccentricity migration for hot and warm Jupiters. We are particularly motivated by the upcoming European Space Agency's Ariel mission that promises to deliver a uniform sample of gas giants over a range of parameter space, ideally suited to identifying trends rather than a deep dive into individual or a small selection of planets.\footnote{While we benchmark all our calculations and discussions to the projected capabilities of Ariel, our results will be broadly applicable for any survey design comparable to Ariel.} Our investigation is therefore complementary to ongoing James Webb Space Telescope (JWST) programs such as BOWIE-ALIGN that specifically targets a handful (8) of hot Jupiters around stars above the Kraft break \citep{Kirk24} and JWST Cycle 4 GO 9025 that specifically targets 5 warm Jupiters that are spin-orbit aligned \citep{Gao-JWST-C4}.

Ariel is scheduled to launch at the end of the decade to perform a survey of exoplanet spectra \citep{2018ExA....46..135T, tinetti2021arielenablingplanetaryscience}. Much of Ariel's four year primary mission will be devoted to transit spectroscopy of hundreds of diverse exoplanets---primarily gas giants \citep{tinetti2021arielenablingplanetaryscience}. As outlined in the Ariel Definition Study Report \citep{2022EPSC...16.1114T}, Ariel would perform observations of four overlapping target lists: Reconnaissance Survey (Tier 1), Deep Survey (Tier 2), Benchmark Planets (Tier 3), and Phase Curves (Tier 4). We focus on the Tier 2 survey, which is expected to make up the bulk of the Ariel transit spectroscopy.

The spectral coverage and signal-to-noise of Ariel Tier 2 transit spectra are sufficient to usefully constrain the abundance of the main oxygen- and carbon-bearing species \citep{2022ExA....53..447B}, hence providing a measurement of atmospheric metallicity \citep{2024SSRv..220...61S}. Tier 2  spectra have planetary signal-to-noise ratios of at least 7 at a spectral resolution of $R = 10$, 50,
and 15 in NIRSpec, AIRS-CH0, and AIRS-CH1, respectively \citep{2021AJ....162..288M}. Uniform observations of this quality for hundreds of exoplanets and their host stars will provide an unprecedented opportunity to test planet formation theory by quantifying trends between atmospheric metallicity, orbital and planetary parameters \citep{2022ExA....53..473D,2022ExA....53..225T}.

Our paper is organized as follows. We outline the underlying theoretical calculations for post-formation pollution in Section \ref{sec:theories}. Resulting trends of atmospheric metallicity and water abundance with respect to orbital period are presented in Section \ref{results}. Methods and the outcomes of simulated Ariel transit survey are shown in Section \ref{sec:ariel}. We discuss shortcomings, implications, and suggestions for optimal observation strategy in Section \ref{sec:discussion}, and conclude in Section \ref{sec:concl}.

\section{Theories}
\label{sec:theories}

We first construct theoretically expected trends in atmospheric metallicity, focusing on the deviation of the planetary metallicity from the stellar value due to post-formation pollution. 
Our assumptions are: 1) the giants inherit the stellar metallicity at formation; and 2) the infalling material is well-mixed with the envelope by convection. 
Following these assumptions, we consider observable atmospheric metallicity as the mass of solids accreted divided by the mass of the planet, fixed to one Jupiter mass for simplicity.\footnote{We have implicitly assumed the mass of the initial rocky core that triggered the nucleation of a giant is significantly smaller than one Jupiter mass, which is valid by the definition of a gas giant.}

\subsection{Pollution by Solid Accretion}\label{mass accretion}

We consider three different paths of pollution: in situ planetesimal accretion, in situ pebble accretion, and migration. In reality, the modes of accretion are likely mixed, but we keep the three paths distinct for clarity. In all cases, we consider a giant planet of mass $M_p = 1M_{\rm jup}$ and radius $R_p = 1R_{\rm jup}$ to have already formed and be simply further accreting the remaining solids in the disk. Our results can be easily scaled to planets of different masses whose implication we discuss in Section \ref{sec:discussion}.

Under planetesimal accretion, we adopt
\begin{equation}\label{Mdot pla}
    \dot{M}_{\rm pla} = \Sigma\Omega R_p^2
\end{equation}
where $\Sigma$ is the local solid disk surface density, $\Omega \equiv \sqrt{GM_\star/a^3}$ is the Keplerian orbital frequency, $G$ is the gravitational constant, $M_\star$ is the mass of the star, $a$ is the orbital distance, and $R_p$ is the radius of the planet. 
We do not consider gravitational focusing as it is unclear a priori what the typical random velocity of the planetesimals should be. We therefore consider the simplest case of geometric cross section, which is a conservative limit.

Next, we consider pebble accretion,
whereby smaller solids that are marginally aerodynamically coupled to the gas are accreted to a planet through gas drag \citep[e.g.,][]{Ormel2010TheProtoplanets}.
This channel requires the preexistence of a massive seed with significant gravity to kickstart and sustain the process.
The rate of pebble accretion depends on the disk gas properties so we first introduce the disk surface density $\Sigma_{\rm gas}$ and the temperature profiles we adopt.
For our fiducial surface density, we adopt the minimum mass extrasolar nebula reported by \citet{Chiang2013}:
\begin{equation} \label{eq:sigma_gas}
    \Sigma_{\rm gas}=1.3\times10^{-5}\left(\frac{a}{0.2\,\rm AU}\right)^{-1.6}{\rm g\,cm}^{-2},
\end{equation}
and for our radial temperature profile, we use 
\begin{equation}
    T = 150\,{\rm K}\,\left(\frac{a}{1\,{\rm AU}}\right)^{-3/7}
    \label{eq:Tdisk}
\end{equation}
from \citet{Chiang97} for a passive disk.

Since the solid accretion occurs onto an already formed Jupiter, pebble accretion is always in the shear regime such that
\begin{equation}\label{Mdot peb}
    \dot{M}_{\rm peb} = \Sigma\Omega\left[ \sqrt{2}R_{\rm Hill}St^{1/3}\right]^2,
\end{equation}
where $R_{\rm Hill} \equiv a \,(M_p/3M_\star)^{1/3}$ is the Hill radius, $St\equiv t_{\rm stop}\Omega$ is the Stokes number, and $t_{\rm stop}$ is the aerodynamic stopping time. We parametrize the size of the particles by varying $St$ from 0.0001 to 0.01 and verify that in all the cases, our accreting planet remains in the shear-dominated regime.

Finally, we consider pollution by migration. For a planet moving radially at speed $\dot{a}$, we can write the rate of mass accretion as 
\begin{equation} \label{Mdot mig}
    \dot{M}_{\rm mig}=2\pi\dot{a}\Sigma R_p,
\end{equation}
where we assume all the material that directly hit the planet as the latter migrates undergo inelastic collision.

As the orbiting planet tidally perturbs the gas, the gas torques back on the planet. Any non-zero net torque will cause the planet to migrate inward or outward, depending on the sign of the net torque. The exact behavior (both the direction and the magnitude) of the migration depends very sensitively on the disk structure, including its turbulent and radiative properties \citep[e.g.,][]{PPVII-disk-planet}. Here, we adopt Type I migration modified for gap opening presented by \citet{Kanagawa18} which likely places an upper limit on the pollution rate by migration. The torque $\tau$ is related to the orbital angular momentum of the planet $L$ via
\begin{equation} \label{adot}
    \tau=\frac{\partial L}{\partial a}\dot{a}=\left(\frac{M_{p}\Omega a}{2}\right)\dot{a}.
\end{equation}
Meanwhile, the net torque from planet-disk interaction is \citep[e.g.,][]{Kley2012Planet-DiskEvolution}
\begin{equation} \label{torque}
    \tau=-\frac{G^2M_{p}^2}{c_s^2}\frac{\Sigma_{\rm gap}}{H}H,
\end{equation}
where $c_s \equiv (kT/\mu m_H)^{1/2}$ is the sound speed, $\mu \equiv2.7$ is the mean molecular weight, $H\equiv c_s/\Omega$ is the disk scale height, and $\Sigma_{\rm gap}$ is the surface density within the gap carved out by the planet,
\begin{equation}
    \Sigma_{\rm gap} \equiv \frac{\Sigma_{\rm gas}}{1+0.034K},
\end{equation}
where
\begin{equation}
    K\equiv\left(\frac{H}{a}\right)^{-5}\left(\frac{M_{p}}{M_{\star}}\right)^{2}\alpha^{-1}
\end{equation}
with $\alpha = 10^{-3}$, the Shakura-Sunyaev turbulent alpha.\footnote{While the $\alpha$ is constrained to be low in the outer regions of the disk ($\lesssim 10^{-4}$; e.g., \citealt{Pinte2015DustRatio}, \citealt{Flaherty2017ADCO+}, \citealt{Lee24-dust}), it is unclear whether $\alpha$ would stay as low at short orbital periods that are relevant to Ariel.} In writing equation \ref{torque}, we ignored order unity numerical coefficient ($\sim$2.5 for our choice of $\Sigma_{\rm gas}$ and $T$). We discuss below in Section \ref{results} how accounting for this coefficient would not change our final result.
Planets were initially placed between 0.5 and 50 AU and we evolve their inward migration by integrating equation \ref{adot}.

We further account for the dissipation of the disk gas using
\begin{equation} \label{sigma gas wrt t}
    \Sigma_{\rm gas}(t)=f_{\rm dep}\Sigma_{\rm gas,0}\,e^{t/t_{\rm disk}},
\end{equation}
where $\Sigma_{\rm gas,0}$ is set by equation \ref{eq:sigma_gas}, $f_{\rm dep}$ is a depletion factor we vary between 0.001 and 1, and $t_{\rm disk} = 2$ Myr is the disk lifetime \citep[e.g.,][]{Mamajek2009InitialDisks}.\footnote{While the more improved disk lifetime is closer to $\sim$7--8 Myr \citep[e.g.,][]{Michel2021BridgingDissipation}, Jupiters are more likely to appear around more massive stars which tend to have shorter-lived disks \citep[e.g.,][]{Ribas15} so we use 2 Myr as our nominal value.}

\subsection{Solid Disk Density Profiles}

We now establish the solid disk surface density profile $\Sigma$. 
Under the minimum mass extrasolar nebula (MMEN) constructed by \cite{Dai2020}, 
\begin{equation}
    \Sigma =\Sigma_0\left(\frac{a}{\rm AU}\right)^{-1.76}
    \label{eq:mmen}
\end{equation}
where $\Sigma_0=50\,\rm{g\,cm^{-2}}$ which is very close to, but nevertheless an update of the result of \citet{Chiang2013}.

\begin{figure}
    \centering
    \includegraphics[width=\linewidth]{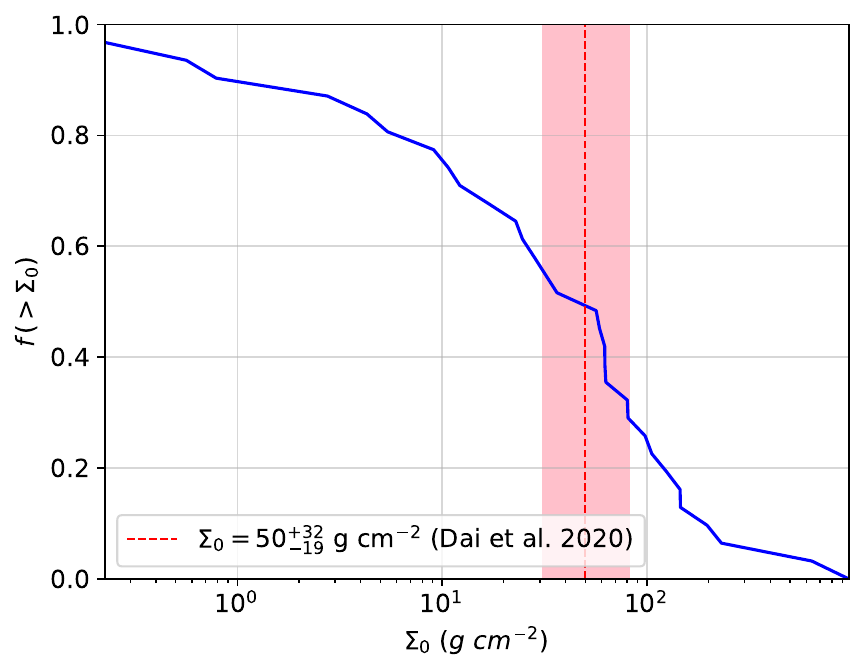}
    \caption{
    Cumulative distribution function of solid disk densities at 1 AU ($\Sigma_0$) calculated from protoplanetary disks in \cite{PPVII} around stars of mass $\geq$1$M_\odot$.} 
    \label{fig:Sigma CDF}
\end{figure}

There is a significant scatter in the normalization of MMEN ($50^{+33}_{-20}\,{\rm g\,cm^{-2}}$) due to the scatter in the exoplanet radius/mass measurements. Likewise, there are orders of magnitude scatter in the measured dust masses in protoplanetary disks, both in more evolved T Tauri phases \citep[e.g.,][]{PPVII} and in young Class 0/I phases \citep[e.g.,][]{Tobin20}. To account for this inter-system variation, we draw $\Sigma_0$ from the dust masses reported in \citet{PPVII}, multiplied by a factor of 3 to obtain the initial solid mass (this multiplicative adjustment brings the T Tauri disk mass into agreement with that of Class 0/I; see, e.g., \citealt{Chachan2023}). The normalization of $\Sigma$ anchored to 1 AU was derived by assuming the disk follows $\Sigma \propto a^{-1.76}$ from 0.1 to 100 AU. 
Figure \ref{fig:Sigma CDF} illustrates the cumulative distribution function of $\Sigma_0$. We see that the result of \citet{Dai2020} corresponds to $\sim$50$^{\rm th}$ percentile, consistent with the idea that MMEN represents a {\it typical} disk that eventually forms exoplanetary systems. 
We note that only the dust masses around stars more massive than 1 $M_\odot$ are chosen for this exercise given that giants are more common around such stars \citep[e.g.,][]{Cumming2008,Fulton2021}.

Under in situ planetesimal and pebble accretion, we account for the corresponding reduction in $\Sigma$ following
\begin{equation}
    \dot{\Sigma}=-\frac{\dot{M}}{2\pi a\delta a}
\end{equation}
where $\dot{M}$ is the mass accretion rate of the planet as described in Section \ref{mass accretion} and $\delta a$ is the width of the feeding zone. Following the arguments outlined in Section \ref{mass accretion},
\begin{equation}
    \left.\delta a\right|_{\rm pla} = R_p
\end{equation}
for planetesimal accretion, and
\begin{equation}
    \left.\delta a\right|_{\rm peb} = \sqrt{2}aSt^{1/3}\left(\frac{M_{p}}{3M_{\odot}}\right)^{1/3}
\end{equation}
for pebble accretion. Both $\dot{M}$ and $\dot{\Sigma}$ are solved simultaneously using the forward Euler method up to 1 and 10 Myr using 32 000 and 320 000 linearly spaced timesteps respectively. Integration is stopped prematurely if the total amount of solid remaining is less than that of a single planetesimal/pebble:
a planetesimal 
is taken to be 4$\times$10$^{12}$g and have a 10 km diameter while a pebble is taken to be 4$\times$10$^{-6}$g and 1 cm across. Under migration, we do not evolve $\Sigma$ but simply stop the planet migration and accretion at 10 Myr or when the planet reaches 10 days---which is the typical expected disk truncation edge under magnetospheric truncation \citep[see, e.g.,][]{Lee17,Batygin23}---whichever comes first.

\subsubsection{Dust Replenishment}

Our model setup implicitly treats the planet feeding zone to be isolated (i.e., no replenishment of solids). For planetesimal accretion, unless there are neighboring planets that can systematically scatter the planetesimals in, there is no a priori reason why a given feeding zone would be replenished. For pebble accretion, the picture is more complicated as the pebbles drift in aerodynamically and the drift timescale is shorter than pebble accretion time by factors of up to $\sim$10 \citep[e.g.,][]{Lin2018,Chachan2023}. One way to stop this radial drift-in is to establish a pressure trap \citep[e.g.,][]{Pinilla2011TrappingDisks}. While the exact source of the pressure trap remains debated---the most popular being perturbation by an embedded planet---it is a natural mechanism to explain both the longevity of dust disks and the observed concentric ring-like structures in protoplanetary disks \citep[see, e.g.,][for a review]{Bae23-PPVII}.

The efficiency with which the dust remains trapped in these pressure bumps has recently been called into question. Early analytic and numerical simulations found the traps to be imperfect only for particles of sufficiently small Stokes number which can be readily filtered through \citep[e.g.,][]{Zhu12}. More recent 2D local shearing box \citep[e.g.,][]{Lee22} and 3D global disk simulations \citep[e.g.,][]{Huang25} report the traps to be leaky even for Stokes number as large as 0.1 when the traps are established by an embedded planet, with the traps becoming more leaky for a lighter planet and larger turbulent diffusion coefficient. Such numerical findings are consistent with recent observations that find the abundance of volatiles in the inner disk to be largely unaffected by the presence of outer substructures \citep[e.g.,][]{Perotti23,gasman25}.

Given the range of possibilities, we explore the two limiting cases. Our isolated in situ pebble accretion corresponds to the case of a perfect trap created by the accreting Jupiter. We introduce another limiting case of a trap that is so inefficient that the trap is functionally non-existent \citep[see e.g.,][for a more sophisticated disk model under this assumption]{Bitsch23}. In this latter case, the planet has access to all the initial dust material exterior to its orbit: 
\begin{equation}
    M_{\rm dust}(>a_{\rm pl}) \equiv 2\pi\int^{100}_{a_{\rm pl}/{\rm AU}}\Sigma_0 \left(\frac{a}{\rm AU}\right)^{-0.76}d\left(\frac{a}{\rm AU}\right)
    \label{eq:Mdust_a_pl}
\end{equation}
where $a_{\rm pl}$ is the orbital distance of the accreting planet. We compute the amount of solids accreted onto the planet by multiplying $M_{\rm dust}(>a_{\rm pl})$ by $\epsilon=0.1$ where $\epsilon$ is the pebble accretion efficiency defined as the ratio of the pebble accretion rate to the radial drift rate \citep[see][for detail]{Chachan2023}. 

\subsubsection{Water Abundance}
\label{sssec:water}

So far, we have computed the total mass of the metals that would be accreted onto a giant planet through post-formation pollution. Creating a large uniform sample of atmospheric measurements is most amenable for volatile species such as water which is more easily detectable than tracers of refractory species such as sulfur or nitrogen. Treatment of water abundance depends on the location at which the solids were accreted. Inside the iceline, water is in gaseous form, so post-formation accretion of solids would not impact water abundance, hence the water abundance is set by the initial solar value, which we take as 0.48 (oxygen mass fraction among all metals; \citealt{GN93}) times 0.02 (our adopted solar metallicity), irrespective of pollution parameters. Outside the iceline, water is in solid form, so we estimate the water abundance by multiplying the total accreted metal mass by 0.48, assuming all the oxygen to be incorporated into water for simplicity \citep[see, e.g.,][their Figure 3, ``cold'' case]{Heng16}. We take the water iceline at 160 K which is $\sim$0.86 AU ($\sim$291 days) based on our adopted disk temperature profile (equation \ref{eq:Tdisk}). For more direct comparison with observable, we further multiply our mass ratio by 2.3/18 (solar metallicity gas mean molecular weight / mass of water molecule) to obtain number ratio. We focus solely on water abundance rather than C/O ratio because proper prediction of the latter requires careful accounting of carbon bearing species as well \citep[e.g.,][]{Mordasini16,Vlahos24} which is beyond the scope of this paper.

\begin{figure}
    \centering
    \includegraphics[width=0.5\textwidth]{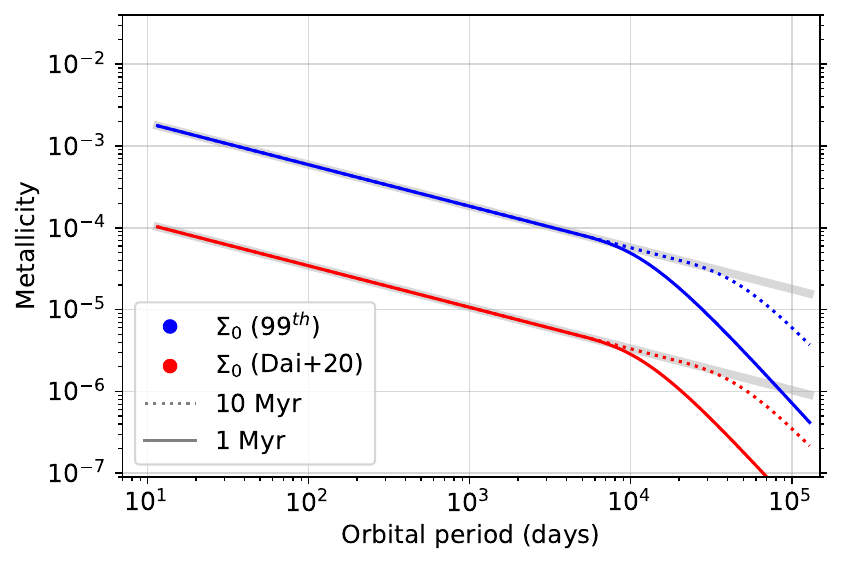}
    \caption{Predicted trend of metallicity with orbital period under local planetesimal accretion. The blue lines corresponds to highest $\Sigma_0$ (99th percentile) while the red lines illustrate $\Sigma_0$ of \citet{Dai2020}. Solid lines show 1 Myr evolution while dotted lines show 10 Myr evolution. The gray lines signify the total available mass budget within the feeding zone.}
    \label{fig:planetesimal-results}
\end{figure}

\begin{figure}
\gridline{\fig{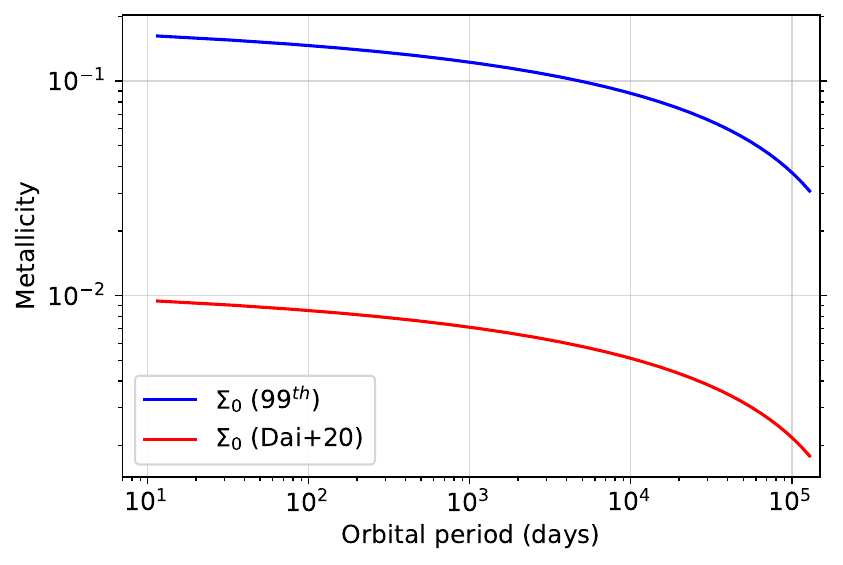}{0.5\textwidth}{(a) No trapping of dust}}
\vspace{-0.5cm}
\gridline{\fig{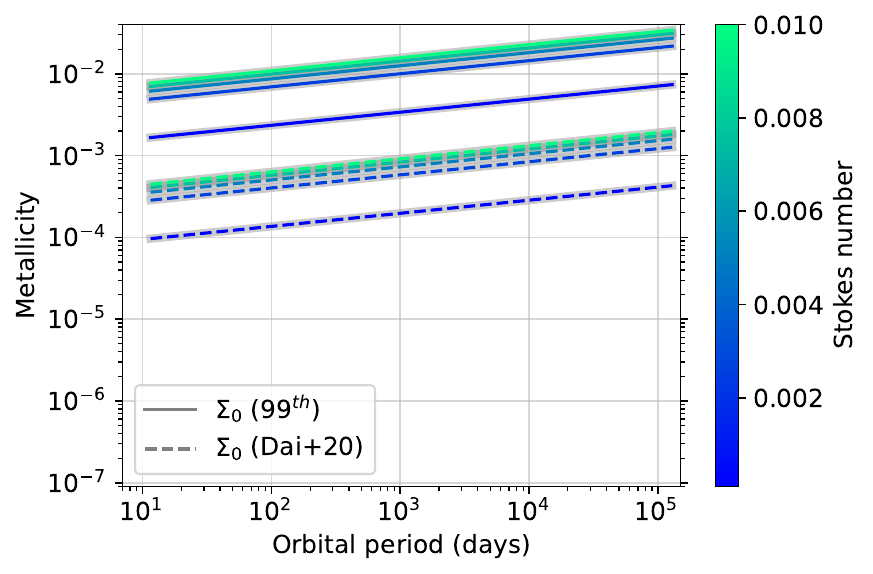}{0.5\textwidth}{(b) Perfect trapping of dust}}
    \caption{Same as Figure \ref{fig:planetesimal-results} but for pebble accretion. The top panel corresponds to no dust trapping while the bottom panel corresponds to perfect dust traps (i.e., local accretion) where different colors correspond to varying Stokes number.}
    \label{fig:pebble-results}
\end{figure}

\begin{figure}
    \centering
    \includegraphics[width=0.5\textwidth]{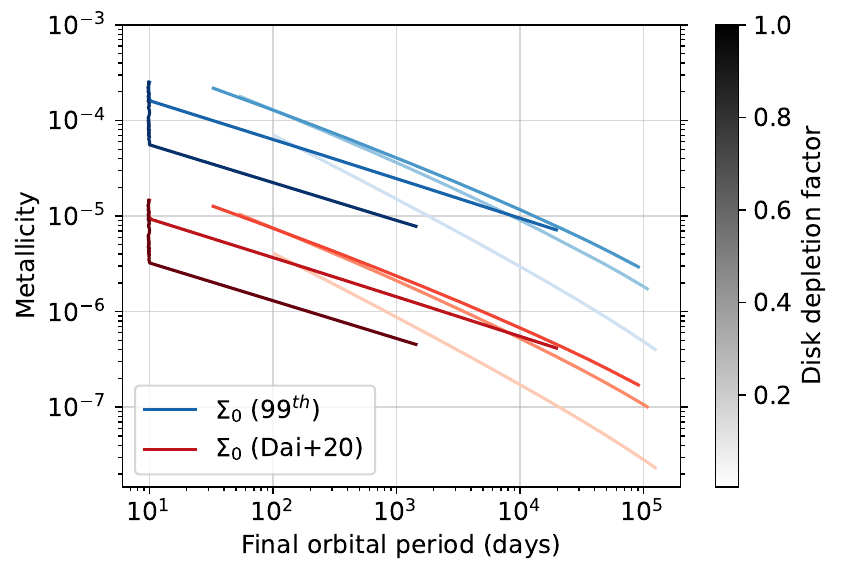}
    \caption{Predicted atmospheric metallicity under pollution by disk-induced migration with respect to the final orbital period of planets. Blue lines correspond to massive disks with $\Sigma_0$ at the 99th percentile while red lines correspond to that of typical MMEN \citep{Dai2020}. Varying levels of transparency correspond to $f_{\rm dep}$ with 1 being no gas depletion. Planets are halted either after 10 Myr or once they reach 10 days which we set to be the inner edge of the protoplanetary disk. Even the highest metallicity falls below the assumed stellar value 0.02 by roughly two orders of magnitude, much more than the order unity numerical coefficient we ignored in the torque calculation.}
    \label{fig: Migration accretion}
\end{figure}

\section{Results}\label{results}

The predicted trends of metallicity (accreted solid mass over Jupiter mass) following post-formation pollution by planetesimal accretion, pebble accretion, and disk-induced migration are found in Figures \ref{fig:planetesimal-results}, \ref{fig:pebble-results}, and \ref{fig: Migration accretion} respectively.
The abundances shown in these figures represent what is added to the planet after formation; therefore, they do not represent the absolute atmospheric abundances---in order to obtain the absolute abundances, our results would have to be added to the metallicities the planets inherited at formation from the disk, which we assume to be stellar for simplicity. The implication is that any additive metallicity that falls below the stellar value (fixed to be 0.02 for simplicity) at a level that is smaller than what is detectable, would simply not show up in observations as a signature of pollution. Since both planetesimal accretion and migration produce additive metallicity that is an order of magnitude smaller than 0.02, we can reasonably rule them out as observable (even if we account for the order unity factor we ignored in the migration rate in case of pollution by migration). Instead, we focus on pebble accretion with high $\Sigma_0$ and high $St\sim0.01$ which produce sufficiently high additive metallicity. We have adopted a conservative case of planetesimal accretion here given the uncertainties introduced by the unknown dynamics of planetesimals---we revisit in Section \ref{sec:discussion} a case for a more optimistic scenario.

Our calculation shows that the metallicity-period trend behaves in a different manner between pebble accretion with and without dust traps with the former declining and the latter rising with orbital period. In the absence of a functional dust trap, a gas giant has access to all the solid material exterior to the planet's orbit. Since we place a hard limit on the size of the disk at 100 AU, if the planet is located farther away from the star, there will be less mass available exterior to it. In the case where a perfect dust trap is established, a giant has access to material only within its feeding zone whose area is $2\pi a\delta a$ and the total mass within this annulus under pebble accretion is
\begin{align}
    M_{\rm peb} &\sim \Sigma_0 \left(\frac{a}{1\,{\rm AU}}\right)^{-1.76}\,2\pi a\left[\sqrt{2}aSt^{1/3}\left(\frac{M_{pl}}{3M_{\odot}}\right)^{1/3}\right] \nonumber \\
    &\propto a^{0.24},
    \label{eq:Mpeb-perf-trap}
\end{align}
which rises at larger orbital distances. The above equation also shows how $M_{\rm peb}$ rises with larger $St$. Conceptually, pebble accretion is more efficient at larger $St$ because aerodynamic drag more easily saps away the momentum of particles of larger $St$ (as long as $St<1$), aiding their gravitational settling onto the planet.

We discuss the metallicity-period trend of planetesimal and disk-induced migration for completeness. Local planetesimal accretion predicts a decline in the level of pollution at wider orbital separations. Repeating the calculation of $M_{\rm peb}$ but under planetesimal accretion whose annular area of feeding zone is $2\pi aR_p$,
\begin{equation}
    M_{\rm pla}=\Sigma_0 \left(\frac{a}{1\,{\rm AU}}\right)^{-1.76}\,2\pi aR_p\propto a^{-0.76}
\end{equation}
which drops with orbital distance. In other words, the key difference is that in {\bf our conservative limit of} planetesimal accretion the planet's accretion cross section stays fixed, whereas in local pebble accretion the accretion cross section is bigger farther from the star.

Likewise, under disk-induced migration, metallicity declines with the final orbital period. By definition, if the final orbital period is long, there has been minimal radial migration and the planet spends most time where $\Sigma$ is lower, leading to overall lower metallicity by pollution. Figure \ref{fig: Migration accretion} shows that the behavior of metallicity with respect to $f_{\rm dep}$ (gas disk depletion factor) is non-monotonic. While the total amount of metals accreted for planets migrating in gas-full ($f_{\rm dep}=1$) disks is larger, these planets also migrate the fastest and reach the shortest final orbital period, causing an overall leftward shift in the metallicity-final-period diagram (e.g., planets that initially start at $\sim$10$^5$ days with the lowest metallicity park at $\sim$10$^3$ days) with the planets that attain highest metallicity all piling up at the disk inner edge of 10 days. 

\begin{figure}
    \centering
    \includegraphics[width=0.5\textwidth]{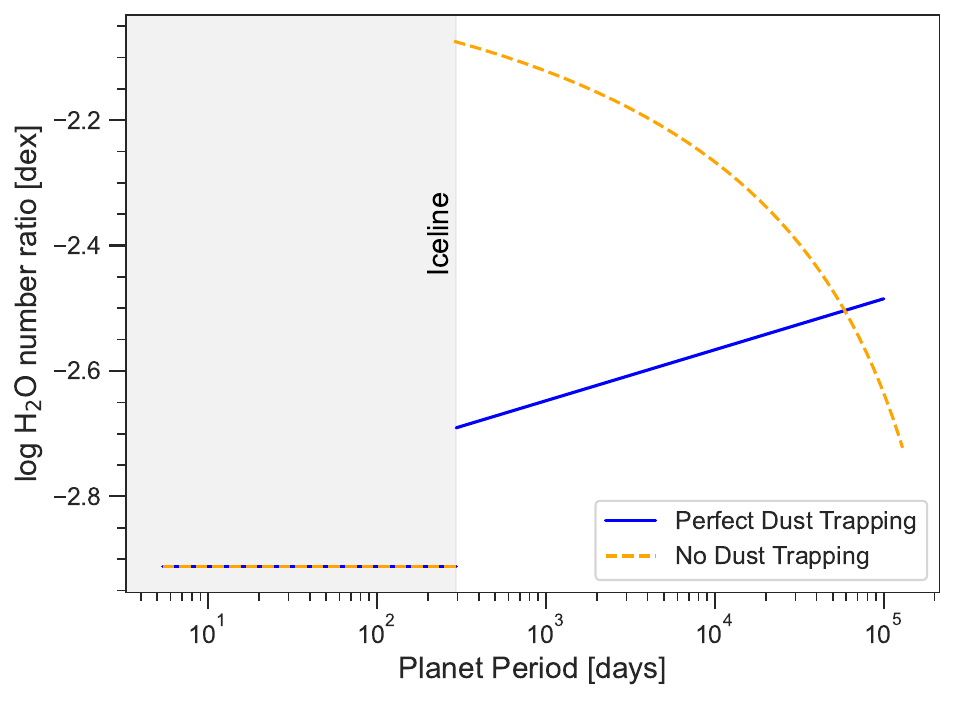}
    \caption{Predicted trend of total water abundance (number ratio) vs.~orbital period. The blue line corresponds to pebble accretion with perfect dust trap with Stokes number = 0.01 and the highest $\Sigma_0$ whereas the yellow line corresponds to pebble accretion with no dust trap at the highest $\Sigma_0$. In both cases, a discontinuity appears at our adopted iceline since inside the iceline, solid pollution does not bring additional water and the total atmospheric water content is that of stellar value.}
    \label{fig:H2O abundance}
\end{figure}

\section{Simulated Ariel Transit Survey}
\label{sec:ariel}

Our aim is to test whether any trends in water abundance with respect to orbital periods that emerge from post-formation pollution can be discerned with an Ariel survey of exoplanet atmospheres. In Figure \ref{fig:H2O abundance}, we illustrate the expected trend of total water abundance with respect to orbital period under pebble accretion, focusing on the two models where pollution is expected to achieve superstellar metallicity. 
In this section, we will use the model-derived water abundance shown in Figure \ref{fig:H2O abundance} to a) quantify how well the atmospheric abundance of a planet could be constrained by Ariel Tier 2 transit spectra, and b) assess whether population-level trends in atmospheric abundance among giant planets can be used to distinguish between different origin channels of hot Jupiters.  

\subsection{Accuracy and Precision of Tier 2 Transit Spectroscopy Retrievals}
\cite{changeat_esa-ariel_2023} simulated Ariel retrievals for Tier 2 transmission spectra, providing estimates of how accurately and precisely water abundance could be measured with such observations. Specifically, atmospheric retrievals were done using \emph{Alfnoor} \citep{changeat_alfnoor_2020} for 21,988 targets, resulting in tables of ground truths and retrieved water abundances, including 16$^{\rm th}$, 50$^{\rm th}$ and 84$^{\rm th}$ percentiles for each target. The median of the retrieval is simply the 50$^{\rm th}$ percentile, while the 1$\sigma$ confidence interval runs from the 16$^{\rm th}$ to the 84$^{\rm th}$ percentile and the 1$\sigma$ uncertainty is half of that range.
An additional cut is performed to only include planets greater than 0.3 Jupiter masses ($M_{\rm jup}$), leaving 6,149 targets.
Using this sample, we find that for planets with water number ratio $>-5$ dex, the expected uncertainty on retrievals from Tier 2 Ariel transmission spectra is $0.10\pm0.07$ dex. 

To simulate Ariel observations using real targets, we use the July 2024 version of the \emph{Ariel Mission Candidate Sample} (MCS) presented by \cite{2022AJ....164...15E} and available on GitHub.\footnote{\bf \url{https://github.com/arielmission-space/Mission_Candidate_Sample}} We use the optimistic version of the MCS, which includes both confirmed and candidate planets.
We conservatively adopt the easiest targets observable in a cumulative observing time of 1 year, where we assume that each transit observation takes 3$\times$ the transit duration and each target is re-observed until it reaches Tier 2 signal-to-noise. The number of observations required for each targets were pulled directly from the MCS. Our mock survey includes 431 Jupiters comprised of 172 confirmed and 259 candidate planets.

\begin{figure}
    \centering
    \includegraphics[width=\linewidth]{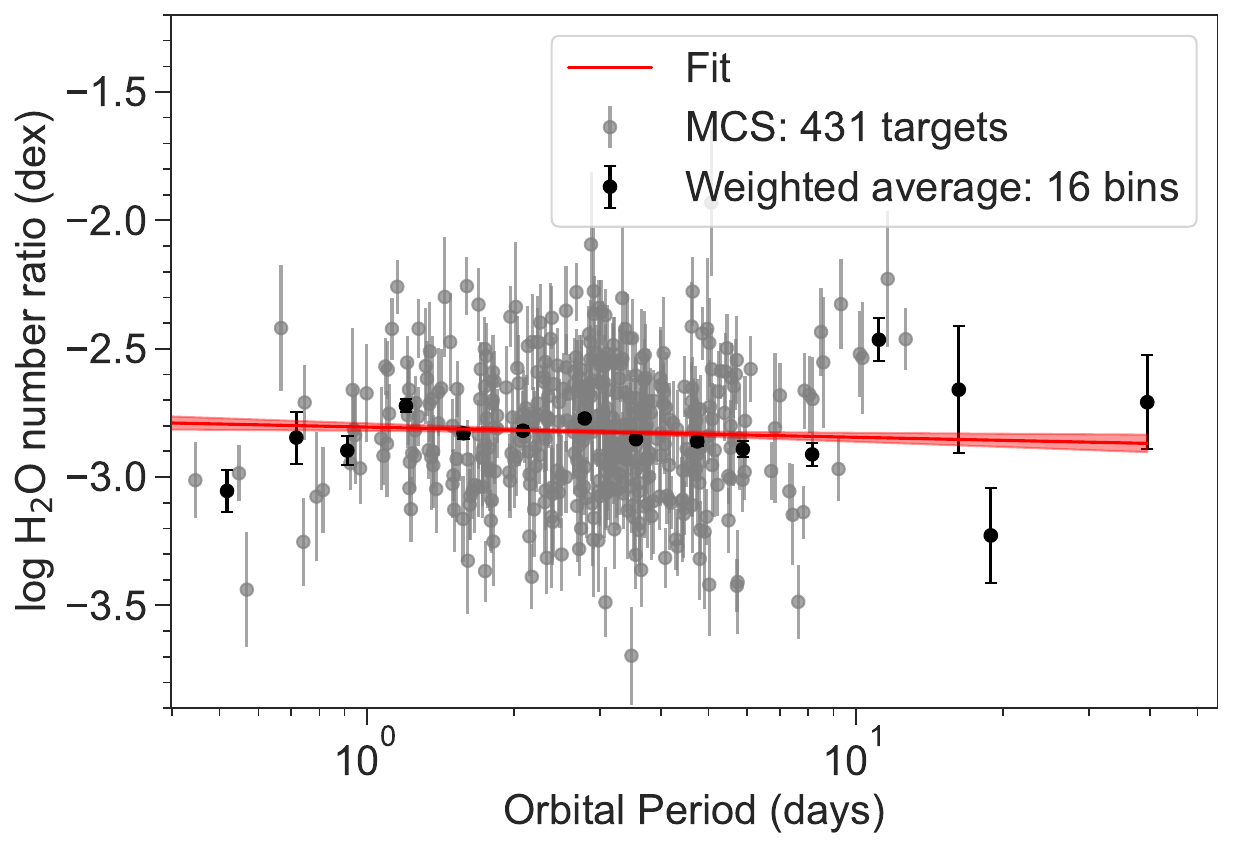}
    \caption{Measured water abundance from simulated Ariel retrievals vs orbital period using targets from the MCS under pebble accretion. 
    The uncertainties on the retrieved water abundance, shown here in gray, are drawn from \cite{changeat_esa-ariel_2023} as described in the text. The transit targets represent the 431 easiest-to-observe Jupiters ($M>0.3M_{\rm jup}$) that can be observed to Tier 2 signal-to-noise ratio in a cumulative observing time of one year. The binned data are shown in black, but the fit shown in red is determined from the unbinned data. The uncertainty on the linear fit is shown by fainter red shading. All targets are within the iceline so a flat H$_2$O abundance with orbital period is expected and obtained regardless of dust trapping. }
    \label{fig: Retrieved metallicity}
\end{figure}

\begin{figure*}
    \gridline{\fig{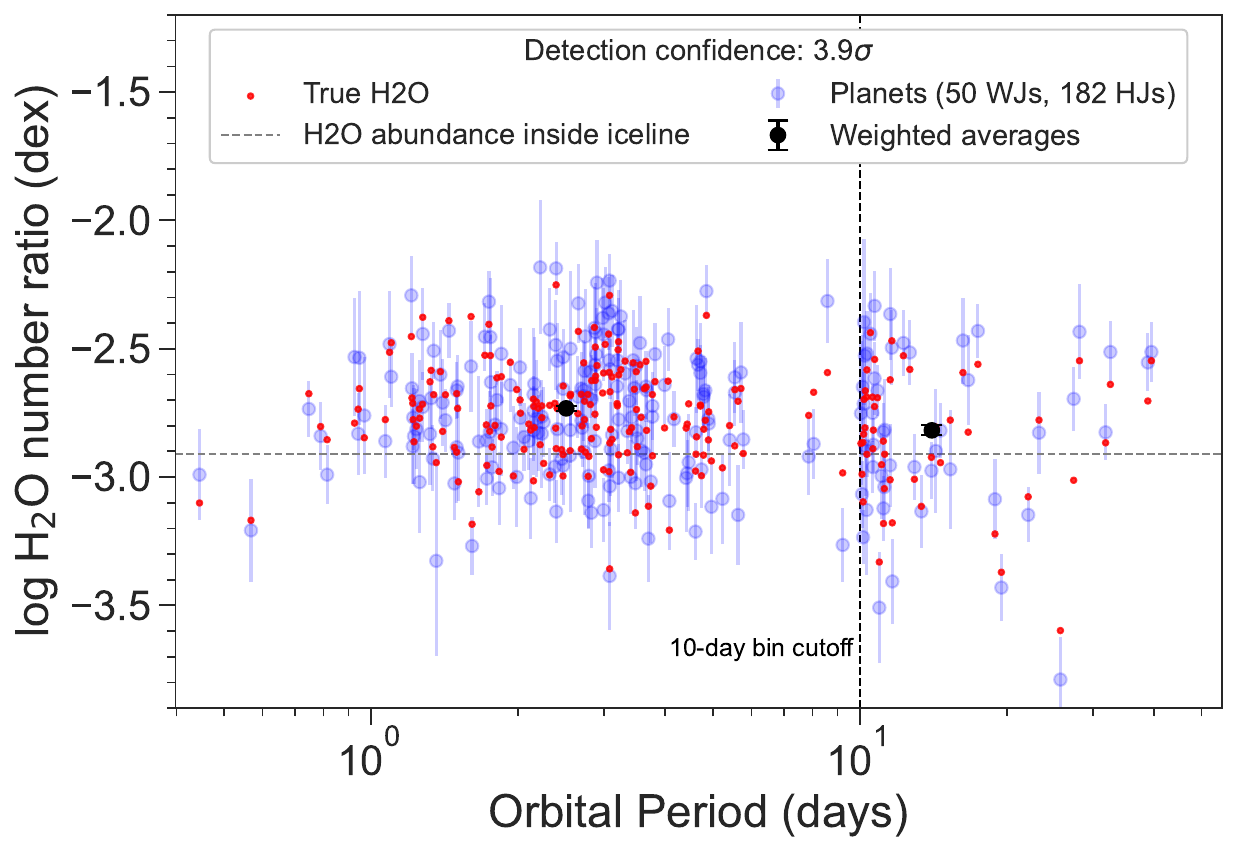}{0.5\textwidth}{(a) Pebble accretion with perfect dust trap}
    \fig{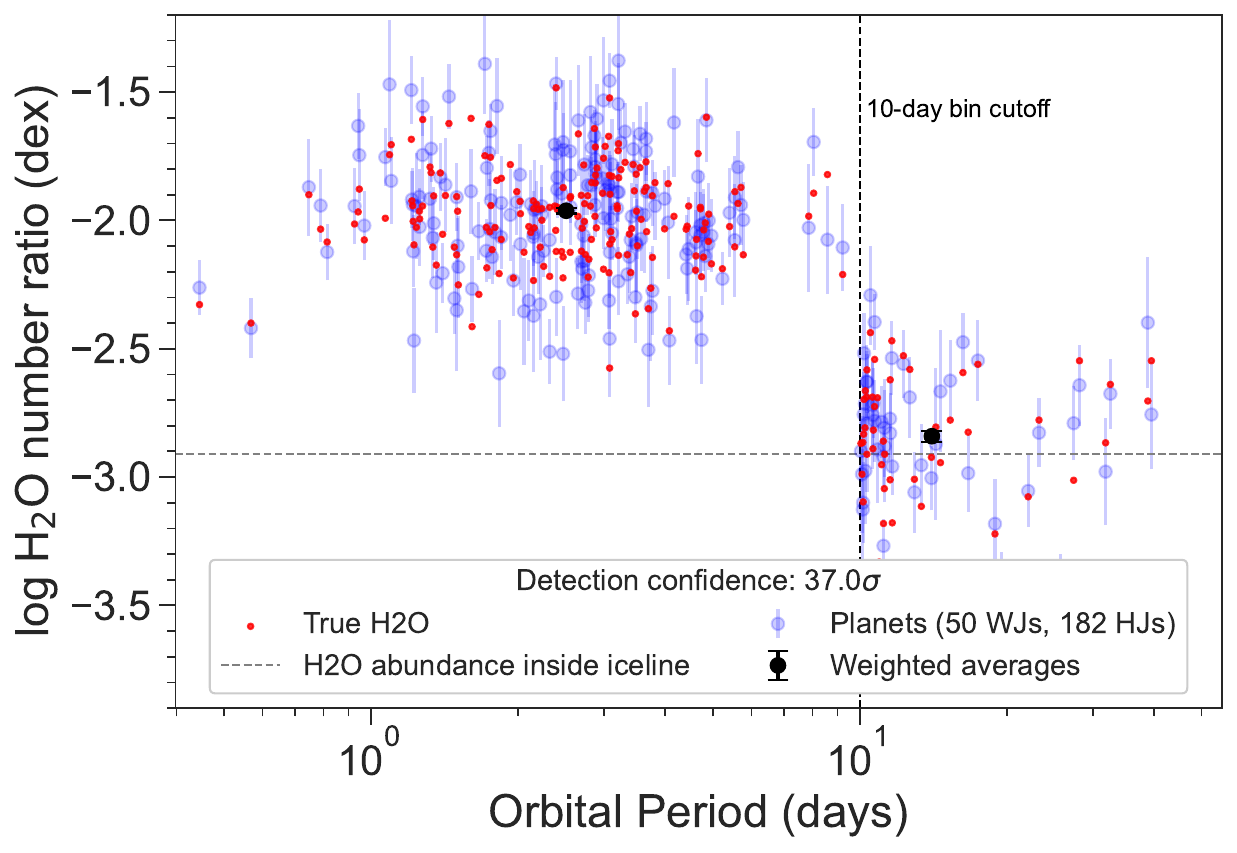}{0.5\textwidth}{(b) Pebble accretion with no dust trap}}
    \caption{Measured water abundance from simulated Ariel retrievals vs.~orbital period using MCS planet data under pebble accretion with perfect dust traps (left) and without a dust trap (right), assuming high-eccentricity migration occurs for Jupiter-mass exoplanets that were originally at $a>1$AU. These are the targets predicted to be observable to Tier 2 within a year. The red points are the assigned true water abundance whereas the blue points are the simulated data accounting for uncertainties. Black points are the weighted average of the hot (left-most point) and warm Jupiters (right-most black point). High-eccentricity migration predicts hot Jupiters to appear significantly more water-enriched than warm Jupiters and the amount of such an excess is expected to be detectable by Ariel.}
    \label{fig: high-eccentricity migration}
\end{figure*}

Ariel's targets are hot and warm Jupiters around FGK and early M type dwarfs so they are all within the iceline. In the simplest baseline scenario of minimal migration after pollution, we expect the water abundance to be flat with respect to orbital period set to the stellar value irrespective of the existence/efficiency of a dust trap, as shown in Figure \ref{fig:H2O abundance}.
We initially assign this stellar water abundance to each planet (adding stellar [Fe/H] to the solar value we adopt) and then randomly shift the abundance by 0.10$\pm$0.07 dex.
Figure \ref{fig: Retrieved metallicity} depicts simulated Ariel retrievals of 431 Tier 2 transmission spectroscopy targets after 1 year of cumulative observation time with Ariel. 
A fit through all the planets (i.e., raw data without binning) recovers the expected flat H$_2$O abundance with respect to orbital period.

\subsection{Disk-Induced vs.~High-Eccentricity Migration}

Under high-eccentricity migration, a gas giant first completes its formation at large orbital separations (beyond the iceline). After the disk dissipates, the giant is excited to a highly eccentric orbit by neighboring planetary or stellar perturbers, then tidally circularizes to a tight orbit with orbital period less than $\sim$10 days.
Under all circumstances, giants that complete their pollution beyond the iceline are expected to have a significantly water-enriched envelope as compared to those that remain inside the iceline (because solid pollution adds extra water only beyond the iceline). We therefore expect the signature of high-eccentricity migration to show up as a sudden jump in water abundance ratio for hot Jupiters compared to warm Jupiters---insofar as most warm Jupiters did not undergo similar dynamical upheaval \citep[e.g.,][]{Rice22-WJ,Wang24}.\footnote{For the purposes of the simulation, we assume that the migrating Jovian planets have already reached an orbital period on the order of days. Additionally, we assume that the metallicity of the planet remains constant throughout the high-eccentricity migration.}

Our goal is to assess the ability of Ariel survey at the Tier 2 level to discern the aforementioned effect of high-eccentricity migration on the measurable water abundance of hot Jupiters. To simulate the migration of a giant from beyond to inside the iceline, 
we assign the water abundance to the MCS targets with orbital periods inside 10 days to that drawn from an arbitrary distance between 1 and 10 AU, from the model expectation in Figure \ref{fig:H2O abundance} corrected to non-solar abundance drawing from stellar [Fe/H] in the MCS, as well as uncertainties drawn from \cite{changeat_esa-ariel_2023} set to 0.10$\pm$0.07 dex. The 10 AU upper limit to the orbital distance is motivated by the observed peak of gas giant occurrence rate at 1--10 AU \citep{Fulton2021}. The MCS targets with orbital periods longer than 10 days are assigned the model-expected water abundance at their measured periods (i.e., stellar value), accounting for the same stellar [Fe/H] correction and measurement uncertainties.
Hot and warm Jupiters are separated, and each sorted to maximize the total number of targets observed. The final list is then a recombination of both, but with a fixed number ratio of 14 hot Jupiters for every 4 warm Jupiters, a ratio representing their relative occurence rate reported by \citet{Petigura18}.

Figure \ref{fig: high-eccentricity migration} illustrates the resulting water abundance of hot vs.~warm Jupiters. As expected, polluted hot Jupiters that undergo high-eccentricity migration from beyond the iceline are expected to be significantly more water-enriched than warm Jupiters. The lower number of Tier 2 targets in this exercise compared to that shown in Figure \ref{fig: Retrieved metallicity} is a result of the additional constraint, wherein we now enforce a minimum number of warm Jupiters which predominantly require more observation time. 

Taking the weighted average between hot vs.~warm Jupiters with the division at 10 days, the detection confidence exceeds $\sim$3--4$\sigma$. In case of no dust trap, the difference is visible by eye. The scenario with no dust traps predicts the hot Jupiters to be noticeably more water-rich than those with perfect dust traps simply because they have access to more solids (see Figure \ref{fig:H2O abundance}). Much of the scatter in true H$_2$O value for hot Jupiters comes from our random distribution of the pre-migration orbital distance without any correlation to the final location in addition to the scatter in stellar metallicity. While a more detailed orbital evolution may beget further trends with e.g., orbital period among the hot Jupiters, we defer such studies to future investigation.

\section{Discussion}
\label{sec:discussion}

\subsection{Stellar and Planet Metallicity}
\label{ssec:metallicity}

The stellar metallicity recorded in MCS draws from the available literature and the NASA Exoplanet Archive, which can be inaccurate. A homogeneous abundance retrieval of $\sim$350 host stars from a subset of potential Ariel targets has been performed by the Ariel Stellar Characterization (ASC) working group (\citealt{ASC25}; see also \citealt{2022ExA....53..473D} for the earlier version). We now check how much the errors in stellar metallicity could impact Ariel's ability to discern the effect of high-eccentricity migration.

\begin{figure}
    \centering
    \includegraphics[width=\linewidth]{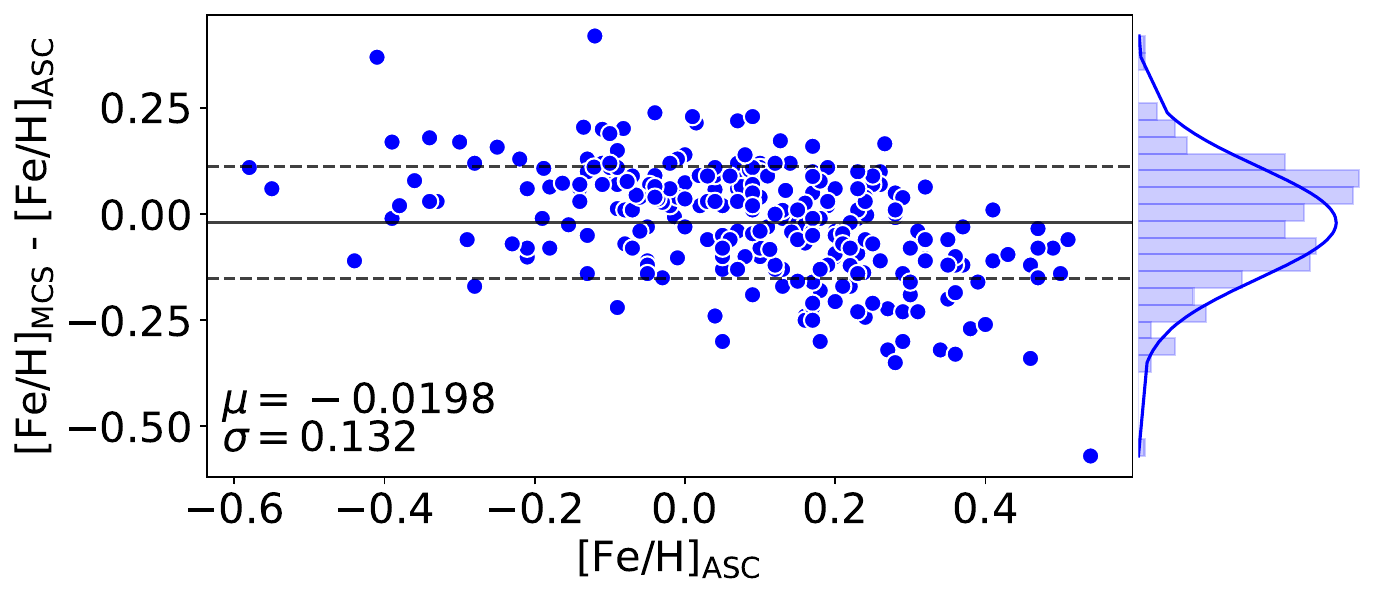}
    \caption{Metallicity comparisons from cross-matched planets in the \textit{Mission Candidate Sample} catalogue \citep[MCS;][]{2022AJ....164...15E} and the Ariel Stellar Characterisation working group's catalogue \citep[ASC;][]{2022ExA....53..473D}.}
    \label{fig: deltaFeH}
\end{figure}

There are 270 host stars with reported metallicities that overlap between ASC \citep{ASC25} and MCS \citep{2022AJ....164...15E}. For these stars, we plot in Figure \ref{fig: deltaFeH} how much the reported metallicities differ.
At subsolar [Fe/H] from ASC, MCS metallicities are systematically larger while at supersolar [Fe/H] from ASC, MCS metallicities are systematically smaller. Such a trend suggests the MCS incorrectly attributes near-solar metallicity to many stars, under-predicting the scatter in measurable H$_2$O abundance in e.g., Figure \ref{fig: high-eccentricity migration}.
The 1-$\sigma$ error that would be introduced by more accurate association of stellar metallicity is 0.132 dex, whose full width ($\pm$1-$\sigma$) is smaller than the target signature of high-eccentricity migration 0.5--1.0 dex. We therefore conclude that the Ariel Tier 2 transit survey remains a viable method of testing the origin theory of hot Jupiters although more accurate measurements of stellar metallicities would be critical to improve both the accuracy at which polluted planets can be identified and the precision to which atmospheric trends can be recovered.

\begin{figure*}
    \centering
    \includegraphics[width=\textwidth]{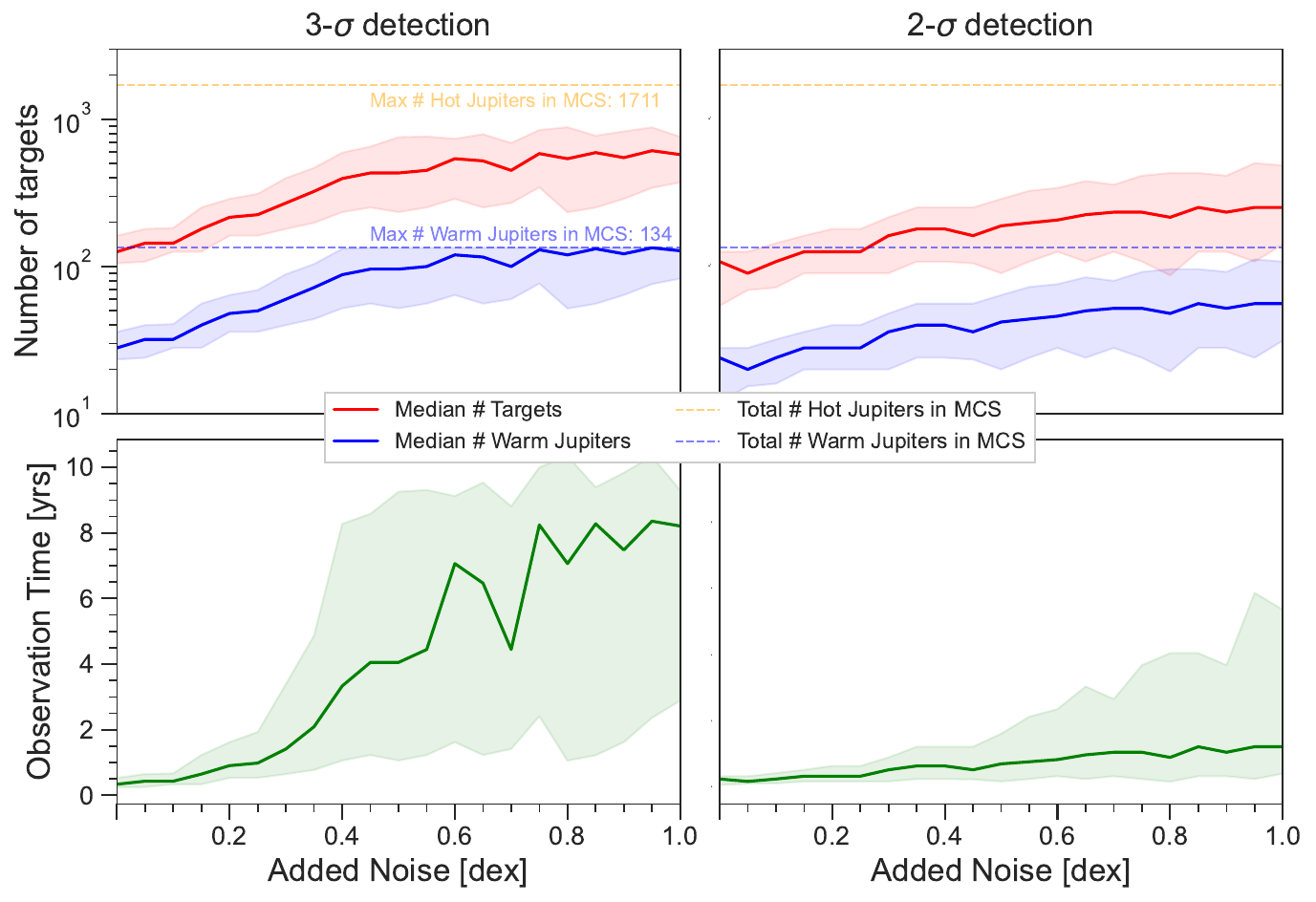}
    \caption{Number of targets required to distinguish the water abundance of hot ($<$10 days) vs.~warm ($>$10 days) Jupiters at 3-$\sigma$ (left column) and 2-$\sigma$ (right column) significance levels as a function of added noise, representing additional scatter in planet metallicity. We focus on the case of pebble accretion with perfect dust trap as the difference in water abundance between hot and warm Jupiters is smaller compared to the case with no dust trap. At a given level of extra noise, we run 100 iterations, reporting the median as the solid line and the 16th and 84th percentiles as the edges of the shaded regions.}
    \label{fig:number-targets-scatter}
\end{figure*}

Further astrophysical scatter can be introduced by any metallicity dependence on planet mass or disk solid mass reservoir \citep[e.g.,][]{Thorngren16}. Under our pebble accretion scenario, the accreted solid mass is much more strongly dependent on the disk solid surface density than the planet mass (equation \ref{eq:Mpeb-perf-trap}) so we expect the scatter introduced by the disk mass to be greater. Alternatively, a scatter in planet metallicity can be achieved by giant impact between gas giants of varying core and gas masses \citep{Ginzburg20}. 
Since the exact source of a mass-metallicity relation or its scatter remains unclear, we compute the required number of targets so that we can still discern the differences in water abundances between hot ($<$10 days) and warm ($>$10 days) Jupiters at $>$2--3 $\sigma$ level for a varying additional scatter in the metallicity data (i.e., in this test, we only have two bins in orbital periods, separated at 10 days).

The result of our analysis for pebble accretion with dust trapping is shown in Figure \ref{fig:number-targets-scatter}. For 2-$\sigma$ detections, the required number of targets is below the 232 expected total number of targets that can reach Tier 2 signal-to-noise within 1 year even if the additional astrophysical scatter is 1.0 dex. In the case of 3-$\sigma$ detections, more targets are required with additional scatter $\gtrsim$0.2 dex. However, with more targets the observation time exceeds the allotted mission time after just $\sim$0.4 dex of additional scatter, since the remaining warm Jupiters require far more time to reach Tier 2. This is a limitation of the dataset, as there are only 134 warm Jupiters with under 50 day periods, many of which are poor Tier 2 transmission targets. Although having more warm Jupiters would inevitably improve the results, it is expected that the final Ariel target sample will be weighted towards the hot Jupiters as they are more easily characterizable and more numerous, perhaps more aggressively than the ratio used here.

\subsection{Formation Condition of Hot Jupiters}

Our results suggest that for post-formation pollution to achieve superstellar atmospheric metallicity (required for observability), the mechanism of pollution must be pebble accretion, the underlying solid disk must be metal-heavy, and in case of perfect dust trap, the Stokes number of pebbles should be $\gtrsim$0.01. The need for metal-heavy disk is consistent with the nucleation of gas giants requiring massive rocky core and the observed correlation between giants and stellar metallicity \citep[e.g.,][]{Fischer2005TheCorrelation}. Pollution by pebble accretion implies that the pollution must occur when the disk gas remains while high Stokes number may correspond to either a reduction in disk gas density---consistent with {\it post}-formation pollution---or larger dust grains. We favor the former explanation based on recent analysis of young protoplanetary disks. By analyzing the shape and the mass content of ringed disks with dust equation of motion, \citet{Lee24-dust} find most ringed protoplanetary disks to feature low Stokes number $\sim$10$^{-4}$--10$^{-3}$. Focusing on IM Lup disk and using detailed disk models, \citet{Ueda24} report that self-consistently explaining multiwavelength observations requires dust grains to be fragile corresponding to similarly low Stokes number.

Classically, planetesimal pollution has been invoked to explain the high atmospheric metallicity of Jovian planets in solar \citep[e.g.,][]{Mousis09} and extrasolar systems \citep[e.g.,][]{Fortney13,Mordasini16}. Our calculations disfavor planetesimal accretion on the basis of the conservative limit of the accretion cross-sectional area presented by the planet's geometric cross section. If the random velocities of the planetesimals can be sufficiently damped below the surface escape velocity of the planets, gravitational focusing will enhance the rate of planetesimal accretion \citep[see][for a review]{Goldreich04} and at its maximal capacity where the accretion radius scales with the Hill radius of the planet, the amount of accreted solid would be comparable to what we expect from pebble accretion with the same scaling of $M_{\rm sol} \propto M_{\rm pl}^{1/3}$. However, planetesimals can also be scattered by planets and planetary embryos which can reduce the rate of accretion with the exact rate dependent on the assumed disk conditions and the size of the embryos and planetesimals \citep[e.g.,][]{Shibata23}, all of which are uncertain.

Alternatively, the final stage of giant formation may be that of giant impact between multiple giants \citep{Ginzburg20}. It is not immediately obvious how the giant impact stage would interface with high-eccentricity migration. The proto-hot-Jupiter may have completed its giant impact beyond the iceline and was scattered inward or it may have undergone a few giant impact on its route towards short orbital period. One way to test this is to select warm Jupiters at eccentricities high enough to be considered proto-hot-Jupiters (i.e., constant angular momentum consistent with a circular orbit at a few day orbital period) and verify whether they have elevated water abundance compared to lower eccentricity warm Jupiters. A similar test can be done between warm Jupiters that are spin-orbit aligned vs.~misaligned \citep[e.g.,][]{Zak25} although orbital misalignment can also realize from local planet-planet scattering (J.~Dong et al.~in prep).

\subsection{Suggestion for Ariel}

Under a uniform period distribution 232 MCS planets, separated into bins of hot and warm Jupiters, we find that Ariel observations at the Tier 2 level of uncertainties are sufficient to test the effect of high-eccentricity migration of hot Jupiters by way of their elevated water abundance ratio compared to warm Jupiters (Figure \ref{fig: high-eccentricity migration}). Even in the relatively pessimistic scenario (perfect dust trap), and with some additional scatter in planet metallicity,
2-$\sigma$ detections are possible with under 252 targets on average, though quite a few more are needed for 3-$\sigma$ detections, in part resulting from a limited number of warm Jupiters in the MCS (Figure \ref{fig:number-targets-scatter}). In other words, the challenge is to sample enough hot \textit{and} warm Jupiters. On the other hand, in the more optimistic case of no dust trapping, less than 100 targets are needed to achieve even $>3$-$\sigma$ detections.

The list of discovered (and confirmed) warm Jupiters by the Transiting Exoplanet Survey Satellite (TESS) continues to grow, so we expect that by the time Ariel launches, there should be enough known warm Jupiters to more definitively test the origin of hot Jupiters, making more viable a formal adoption of the orbital period as one of the axes of diversity in selecting Ariel targets \citep[e.g.,][]{2025arXiv250606429C}.
We therefore suggest attention be given to quantify the likelihood of obtaining Tier 2 level spectra of this growing list of confirmed Jovians beyond 10 day orbital period. 

Our proposed baseline observational test of comparing hot vs.~warm Jupiter water abundance is based on the assumption that warm Jupiters form inside the iceline. Such a test would prefer warm Jupiters that are dynamically cold (i.e., low eccentricity, spin-orbit angle, existence of nearby companions). On the other hand, dynamically hot warm Jupiters (i.e., those with high enough eccentricity to be considered proto-hot-Jupiters) would also be valuable targets: if they also feature elevated water abundance, it would further support the high-eccentricity migration origin of hot Jupiters. We suggest including both dynamically hot and dynamically cold warm Jupiter population in the target sample.  

While a hierarchical model fit \citep[e.g.,][]{2022MNRAS.509..289K} could improve precision, we consider it unnecessary for the test of high-eccentricity migration. 
A more critical requirement is to build a target list that has enough representation of {\it both} 
hot and warm Jupiters (but not necessarily equal representation) for which deep survey (Tier 2) Ariel spectra can be obtained.
Another critical observational effort in this regard is to develop an accurate and uniform set of host star metallicities \citep[e.g.,][]{2022ExA....53..473D,ASC25} to first determine whether a given planet has superstellar metallicities which would be a criterion one can use to identify candidate planets that underwent post-formation pollution.

\subsection{Connection to Cold Jupiters}

Our discussion thus far focused on using the difference in hot Jupiter's water abundance with respect to warm Jupiters under high-eccentricity migration. In our simple baseline scenario of warm Jupiters barely undergoing any post-formation migration, we do not expect to find any trend in water abundance with respect to orbital period among the warm Jupiters. A trend appears only beyond the iceline and Figure \ref{fig:H2O abundance} demonstrates how the behavior of such a trend differs significantly between the case with and without functioning dust traps. While directly resolving the full trend is not possible with current technology, one may compare the hot Jupiter's water abundance ratio from e.g., Ariel with that of directly imaged planets from e.g., James Webb Space Telescope (JWST). Under high-eccentricity migration, hot Jupiters would originate from 1--10 AU ($\sim$365--10$^4$ days) while directly imaged planets would be beyond $\sim$10s of AU. If there was effectively no dust trapping, we expect hot Jupiters to be more water-enriched than wide-orbit planets. If on the other hand there was a perfect dust trap, we expect the opposite. Systematic comparison of atmospheric abundances between hot and cold Jupiters therefore can be used as a test of the efficiencies of dust traps in protoplanetary disks that nucleate gas giants. Such studies would likely require calibrating abundance measurements from transit, emission and reflectance spectroscopy, since the short-period and long-period planets would have to be characterized via different techniques.

\section{Conclusion}
\label{sec:concl}

We explored three post-formation pollution mechanisms---planetesimal accretion, pebble accretion, and disk-induced migration---to construct a theoretical prediction of gas giants' atmospheric water abundance as a function of orbital period. Only pebble accretion in highly metal-heavy disks is able to achieve supersolar metallicity in giant planet atmosphere leaving a detectable signature of pollution with the caveat that a more optimistic treatment of pollution by planetesimal accretion can also achieve supersolar metallicity. If the dominant origin channel of hot Jupiters is high-eccentricity migration while that of warm Jupiters is more dynamically quiescent, we expect a significant upward shift in their atmospheric water abundance compared to warm Jupiters, at a level that is detectable with Ariel. To ensure that it can constrain these formation/migration scenarios, the Ariel Tier 2 transit survey should include $\mathcal{O}(100)$ hot ($<$10 days) {\it and} warm ($>$10 days) Jupiters with at least $\sim$20 warm Jupiters beyond 10 days. Probing both dynamically hot and dynamically cold warm Jupiters is recommended for a more complete test of high-eccentricity migration origin of hot Jupiters. We further suggest a comparison of median hot Jupiter water abundance with that of wide-orbit cold Jupiters which can be leveraged to reveal the property of primordial protoplanetary disks such as the existence and the efficacy of the dust trap established by the embedded giant planet(s).

\begin{acknowledgments}
We thank Quentin Changeat for clarifying discussion on Ariel uncertainties and Daniel Thorngren for useful exchange on mass-metallicity relationship. We also thank Quinn Konopacky, Jean-Baptiste Ruffio, and Jason Wang for useful discussions on atmospheric characterization of directly imaged planets, and James Kirk on short-period giants. An anonymous referee provided an encouraging report that helped improve the manuscript. Geoffrey Bryden, David Ciardi, Mark Swain, and Jiří Žák provided useful feedback.
LD was supported by the Trottier internship from the Trottier Institute for Research on Exoplanets.
BCN gratefully acknowledges support by the Trottier Space Institute and McGill University. 
EJL was supported by NSF Research Grant 2509275, NSERC Discovery Grant RGPIN-2020-07045, DGECR-2020-00230, FRQNT/NSERC NOVA Grant FRQ-NT 2023-NOVA-325929, NSERC ALLRP 577027-22, and the William Dawson Scholarship from McGill University.
NBC acknowledges support from a Canada Research Chair, NSERC Discovery Grant, and McDonald Fellowship.
This work made use of the Ariel Stellar Catalogue developed by the Stellar Characterisation WG in preparation of the ESA Ariel space mission.
\end{acknowledgments}

\begin{contribution}

LD performed the theoretical modelling of expected metallicities.
BCN carried out modelling of observational expectations of Ariel.
EJL conceived and supervised the overall project.
NBC supervised BCN and provided observational and data analysis expertise.

\end{contribution}

\bibliography{ariel-HJ}{}
\bibliographystyle{aasjournalv7}

\end{document}